\documentstyle[twocolumn,aps,epsf,floats]{revtex}
\begin{document}
\twocolumn[\hsize\textwidth\columnwidth\hsize\csname@twocolumnfalse%
\endcsname
\draft
\preprint{IUCM98-xxx}
\title{Electrodynamics of a Clean Vortex Lattice}
\author{W.A. Atkinson and A.H. MacDonald}
\address{Department of Physics, Indiana University, Bloomington IN 47405}
\date{\today}
\maketitle

\begin{abstract}

We report on a microscopic evaluation of electrodynamic response for
the vortex lattice state of a model s-wave superconductor.  Our calculation
accounts self-consistently for both quasiparticle and
order parameter response and establishes the collective 
nature of linear response in the clean limit.
We discuss the effects of homogeneous and inhomogeneous
pinning on the optical conductivity and the penetration depth, and 
comment on the relationship between macroscopic and local penetration
depths.  We find unexpected relationships between pinning arrangements
and conductivity due to the strongly non-local response.

\end{abstract}
\pacs{74.25.Gz,74.60.-w,74.60.Ge,74.25.Nf}
]

The study of vortex electrodynamics in type II superconductors has
been ongoing for over 30 years\cite{review}.  Due to the complexity of
the problem, however, the theory of vortex electrodynamics remains
incomplete.  For a system with perfect translational invariance, it
can be shown rigorously that the long-wavelength response consists
only of a resonance at the cyclotron frequency\cite{drewhsu}, so that
there is no signature of superconductivity in the a.c.\ conductivity.
In real systems the response is complicated by broken translational
symmetries associated with disorder, pinning defects and even the
underlying atomic lattice.  Relatively few rigorous
results exist for the general case.  One regime which has been
extensively studied is the
the dirty limit of vortex core sizes $\xi$ much larger than
quasiparticle mean-free-paths $\ell$.  In this case, the response
is approximately local and the
Bardeen-Stephen\cite{bardeenstephen,gorkov} isolated vortex model 
provides a useful picture.  
Even in the dirty limit, vortex-vortex interactions play an important
role in the electrodynamics except at fields very close to $H_{c1}$ 
and microscopic models rapidly become
intractable.  Instead, one typically interprets experiments at sub-THz
frequencies in terms of simpler phenomenological hydrodynamic
models\cite{gr,cc,revenaz}, in which the mixed state is characterized by
vortex pinning, viscosity, and interaction parameters.

No comparable phenomenology exists for the clean-limit vortex lattice,
for which $\ell$ is larger than $\xi$ or even larger than the distance
between vortices.  In this case the details of disorder within a
particular vortex core\cite{russian} become important and nonlocal
response can make the connection between single-vortex
models\cite{sonin} and the dense vortex systems tenuous.  Interest in
the clean limit has grown in recent years with the advent of clean
single crystals of type II low (s-wave) and high (d-wave) $T_c$
materials.  In particular, the extent to which in-field optical
conductivity measurements can be used as a probe of fundamental
properties of high $T_c$ materials is a subject of great
interest\cite{mallozzi,lemberger}.  Existing experimental work is
confusing and possibly contradictory.  Resonances have been observed
in thin films of YBa$_2$Cu$_3$O$_{7-\delta}$ (YBCO) and were
interpreted as being due to vortex core transitions\cite{drewexpt},
the clean-limit manifestation of vortex motion\cite{hsu1}.  Other,
possibly related, resonances observed in Bi$_2$Sr$_2$CaCu$_2$O$_8$
(BSCCO)\cite{ongexpt} were interpreted as being due to collective
vortex motion.  In contrast, recent experiments\cite{mallozzi}
in thin films of BSCCO were convincingly interpreted in terms of
a continuum of extended quasiparticle states, ignoring vortex motion.  
Finally, infrared measurements of YBCO crystals have failed to find any
appreciable field-dependence to the conductivity\cite{tanner} at all!

Much of the theoretical work on electrodynamics in the clean limit has
been based on one of two extreme pictures; in the first\cite{zhang}
quasiparticle states of the vortex lattice respond directly to the
perturbing electromagnetic field without any order parameter response,
while in the second\cite{russian,hsu1} it is the vortices that move,
usually rigidly, and indirectly disturb the quasiparticle equilibrium.
Another stimulating approach uses a phenomenological composite model to
interpret the multiple resonances and magnetooptical activity seen  
in experiment\cite{drewtheory}.  The difficulty which arises in any
serious calculation, however, is that quasiparticles and vortices are
not separate entities.
A time-dependent order parameter, for example one
arising from vortex motion,  creates a time-dependent potential for
the quasiparticle states from which the order parameter is in turn 
constructed.

In this Letter we report on a microscopic calculation which treats the
order parameter and the quasiparticles on equal footing
\cite{eschrig}, and which naturally captures the nonlocal nature of
clean-limit vortex electrodynamics.  In this case, the appropriate
language is not that of quasiparticles interacting with moving
vortices, but is instead that of collective modes.  The approach is
microscopic and proceeds in three steps.  First we calculate a
self-consistent solution for the equilibrium vortex lattice using the
Bogoliubov-deGennes (BdG) equations.  Second we evaluate the 
linear response of normal and anomolous blocks of the one-particle
density matrix to a perturbing electromagnetic (EM) field,
accounting for vortex motion and distortion self-consistently. 
Third, we use the density matrix to find the induced current,
and therefore the conductivity.   Since this is a linear response 
calculation it addresses external fields which produce 
a vortex displacement that is small compared to the
coherence length.  Because the equations are solved in real space, it
is a trivial matter to incorporate disorder or other forms of broken
translational invariance.

There are two important energy scales in our discussion.  The first is
the cyclotron energy $\hbar \omega_c$, which is the resonant frequency
of a translationally invariant superconducting system \cite{drewhsu}.
The second is the quasiparticle absorption energy $\hbar
\omega_{\mbox{\scriptsize qp}} \sim \Delta^2/W \sim \hbar \omega_c
(H_{c2}/H) $\cite{carolidegennes}, where $\Delta$ is the s-wave gap,
$W$ is the metal band-width, and $H$ is the field-strength. This is
the predicted energy of allowed transitions between vortex core states
in non-self-consistent calculations which ignore vortex motion\cite{zhang}.
Note that the cyclotron frequency drops below the quasiparticle 
bound state transition energy as the external field drops below
$H_{c2}$.
Before turning to a discussion of our results, we briefly outline the
formulation of vortex-lattice equilibrium state and response-function
theory on which they are based.

We have obtained self-consistent solutions of mean-field equations for
equilibrium and linearly perturbed states of two-dimensional
generalized Hubbard model Hamiltonians.  We use a Nambu vector
notation, defining ${\bf C} = (c_1,c_2,\ldots,c_{2N})$ where, $c_i =
c_{i\uparrow}$, $c_{i+N} = c^{\dagger}_{i\downarrow}$,
$c^{\dagger}_{i\sigma}$ and $c_{i\sigma}$ are creation and
annihilation operators for site $i$ and spin $\sigma$, and $N$ is the
number of lattice sites.  The mean field hamiltonian can then be
written as ${\cal H}^{MF} = {\bf C}^{\dagger} {\bf H^{BG}} {\bf C}$
where ${\bf H^{BG}}$ is the $2N \times 2N$ Bogoliubov-de Gennes
matrix:
\begin{equation}
{\mathbf H^{BG}} = \left[ \begin{array}{cc}
{\mathbf \tilde \epsilon} - \mu {\bf 1} & {\mathbf \Delta^0} \\
{\mathbf {\Delta^0}^\dagger} & -{\mathbf \tilde \epsilon^\ast} 
+ \mu {\bf 1} \end{array} \right ].
\label{bdgmatrix}
\end{equation}

Here ${\bf \tilde \epsilon}$, and ${\bf \Delta}$ are $N\times N$
matrices and $\mu$ is the chemical potential.  The one-body matrix
elements, $\tilde \epsilon_{ji}$ can be used to model intersite
hopping, and diagonal or off-diagonal disorder.  The magnetic field,
which we take to be uniform, is accounted for by multiplying all
one-particle matrix elements by Landau gauge phase factors; $\tilde
\epsilon_{ji} = \epsilon_{ji} \exp ( i \alpha Y_i (X_j - X_i) )$ where
$(X_i,Y_i)$ is the location of the $i$-th lattice site and $\alpha =
eB/\hbar c$.  $\Delta_{ij} = V_{ij} \langle c_{j\downarrow}
c_{i\uparrow} \rangle$ where $V_{ij}$ is the Hubbard model interaction
strength between sites $i$ and $j$ and the angle brackets denote a
consistently determined thermal average.  We enforce
periodicity\cite{block,longpaper} in a supercell containing an
even-integer number of vortices, typically two.  The mean-field
Hamiltonian can be written in the diagonal form, ${\cal H}^{MF} =
\sum_n E_n \gamma_n^{\dagger} \gamma_n$ where $E_n$ is an eigenvalue
of $\bf H^{BG}$, $\gamma_n = \sum_i U_{n,i}^\ast c_i$ is a fermionic
quasiparticle annihilation operator, and ${\bf U}$ specifies the
unitary transformation that diagonalizes $\bf H^{BG}$. ($ {\bf
U}^\dagger {\bf H^{BG}} {\bf U} = {\bf E}$ where $\bf E$ is the
diagonal matrix with elements $E_n$.)  The equilibrium density matrix
is $\rho^0 = \exp(-\beta {\cal H})/Z$ with $Z = \mbox{Tr }\exp(-\beta
{\cal H})$, $\beta = 1/k_BT$, and the order parameter equation is
\begin{equation}
\Delta^0_{ij} = V_{ij} [{\bf U} {\bf \rho^0} {\bf U}^\dagger]_{i j+N},
\label{gapfunction}
\end{equation}
Note that in our notation, half of the quasiparticles states are
occupied in the ground state.  Eq.~[\ref{gapfunction}] is the
vortex-lattice state analog of the simple Meissner state gap equation
and can be solved iteratively.

The linear response of the density matrix at frequency $\omega$ is
given by the following familiar expression:
\begin{equation}
\delta \rho_{n',n}(\omega) = \frac{f(E_n) - f(E_{n'})}{\hbar \omega
-E_{n'}+E_{n} + i \eta} [{\bf U}^{\dagger} \delta 
{\bf H^{BG}} {\bf U} ]_{n',n}
\label{linearresponse}
\end{equation}
$\delta {\bf H^{BG}}$ has a contribution ($\delta {\bf \tilde
\epsilon}$) in its diagonal blocks, which describes the coupling of a
weak a.c.\ electric field to the system\cite{longpaper,hartreecaveat},
and a contribution in the off-diagonal blocks,
\begin{equation}
\delta \Delta^\prime_{ij}(\omega) = V_{ij} [{\bf U}
\delta {\bf \rho} (\omega) {\bf U}^\dagger]_{i j+N},
\label{backflow}
\end{equation}
which describes the self-consistent response of the order parameter.
In non-self-consistent calculations the latter contribution is
neglected.  Multiplying, Eq.~(\ref{linearresponse}) by the energy
denominantor, leads to a set of linear equations for $\delta
\rho_{n'n}$ which can be solved by inverting a
symplectic\cite{longpaper} matrix that is singular when $\hbar \omega$
equals a collective excitation energy of the system.  Since
\begin{equation}
\delta O = \sum_{n',n} \langle n' | O | n \rangle \delta \rho_{n,n'},
\label{expectationvalue}
\end{equation}
we can calculate the linear response of any observable to the
perturbing a.c.\ field.  The complex optical conductivity tensor
$\sigma_{\mu \nu}(\omega)$ is defined by $\delta J_\mu = \sum_\nu
\sigma_{\mu \nu}(\omega) E_\nu(\omega)$, where $\bf J$ is the current
operator and ${\bf E}(\omega)$ is the a.c.\ electric field.  Here we
report results for an on-site attractive pairing model ($V_{ij} = -
V_0 \delta_{i,j}$) which leads to s-wave superconductivity.  It is
convenient to compare normal and superconducting state response by
adjusting the strength of the pairing interaction, rather than field
strength or temperature.  The chemical potential is chosen near the
bottom of the band where the dispersion is nearly parabolic.

\begin{figure}[tb]
\begin{center}
\leavevmode
\epsfxsize \columnwidth
\epsffile{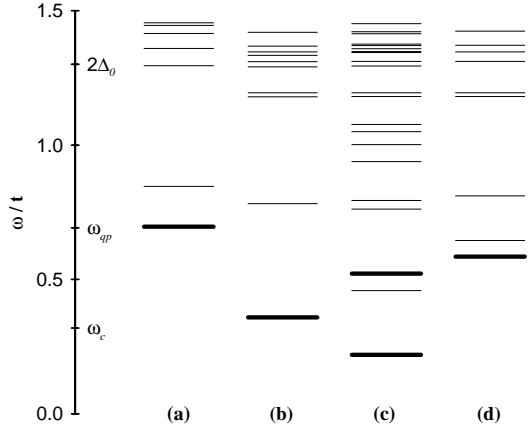}

\caption{Spectrum of the nonequilibrium density matrix
$\delta \rho(\omega)$ at $T=0$.  (a) Non-self-consistent calculation,
(b) Self-consistent calculation, (c) With single pinning center per
unit cell, (d) With two pinning centers per unit cell.  Optically
active resonances with a large spectral weight are indicated by thick solid
lines.   The weaker spectral weights are smaller by at least 
one order of magnitude.  The resonant frequencies in (a) correspond to
quasiparticle energy differences.  Results
are for a magnetic unit cell containing two vortices and 32 lattice
sites.}

\label{fig:one}
\end{center}
\end{figure}

\begin{figure}[tb]
\begin{center}
\leavevmode
\epsfxsize \columnwidth
\epsffile{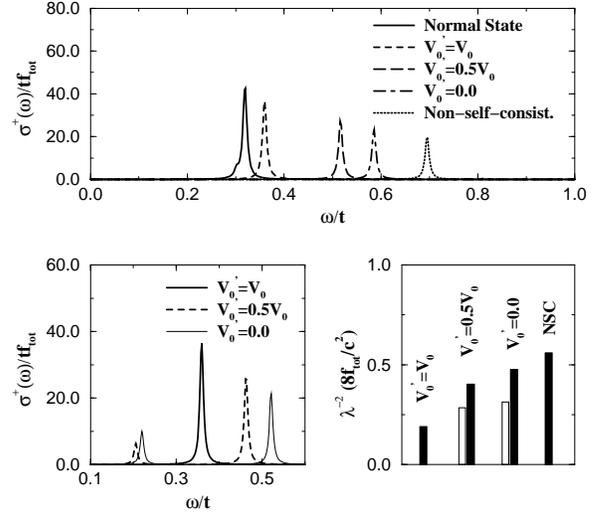}

\caption{Real part of the electron-active circularly polarized
conductivity of vortex lattice states.  
{\em Top:} $V_0^\prime$ has the same value at all
vortex cores.  Results of a non-self-consistent calculation are also
shown.  {\em Bottom left:} Every second vortex core is pinned.  {\em
Bottom right:} Effect of pinning on the penetration depth.  Open bars
are for inhomogeneous pinning.  Conductivity is measured in units of
$tf_{tot}$ where $t$ is the nearest neighbour hopping energy.}

\label{fig:two}
\end{center}
\end{figure}

Resonances in the conductivity are determined by the pole structure of
the nonequilibrium density matrix $\delta \rho(\omega)$, shown in
Fig.~\ref{fig:one}.  While $\delta \rho(\omega)$ has a complicated
spectrum, only a small subset of all possible resonances are excited
by a long wavelength EM field, and the remaining resonances are
optically silent.  In fact, in the absence of pinning and disorder
there is essentially only one optically active resonance, which occurs at a
frequency $\omega_0$ near $\omega_c$, and well below
$\omega_{\mbox{\scriptsize qp}}$, as shown in the top panel of
Fig.~\ref{fig:two} where the real part of the circularly polarized
conductivity [$\sigma^+(\omega) \equiv \sigma_{xx}(\omega) +
i\sigma_{xy}(\omega)$] is plotted.  We stress that the absence of a
peak in the conductivity at $\omega_{\mbox{\scriptsize qp}}$ is not a
product of changes in optical selection rules brought about by vortex
motion, but rather reflects the absence of a pole in $\delta
\rho(\omega)$ at that frequency, as shown explicitely in
Figs.~\ref{fig:one} (a) and (b).  This underscores the collective
nature of the excitation spectrum of the vortex lattice.

In the absence of extrinsic pinning, the small frequency shift
$\omega_0 - \omega_c$ of the response is a result of broken
translational invariance by the lattice.  In general, we expect vortex
motion to be strongly influenced by the atomic lattice when the
coherence length is comparable to the atomic lattice spacing, a
situation which may in fact be realised in BSCCO\cite{fisher}.  The
effects of including homogeneous extrinsic pinning are also shown in
the top panel of Fig.~\ref{fig:two}.  Pinning is modelled here by reducing
the pairing interaction $V_0$ to a smaller value $V_0^\prime$ at each
vortex core center,  a change which has no effect on the equilibrium
vortex lattice state.   Pinning shifts the
resonance to higher frequencies and reduces the 
spectral weight of the mode, with the missing spectral weight
appearing in the zero-frequency (superfluid) response (bottom right
panel of Fig.~\ref{fig:two}).  When $V_0^\prime = 0$, the pinning is
strongest and the resonant frequency is close to
$\omega_{\mbox{\scriptsize qp}}$.  This is consistent with
expectations: as the order parameter response is supressed the
collective mode should increasingly resemble quasiparticle pair creation.
{}From the Hall-angle sum rule\cite{coleman,longpaper}, there is a direct
relation between $\omega_0$ and the spectral weight of the mode or,
equivalently, the penetration depth: 
\begin{equation} 
\lambda^{-2} =
\frac{8f_{tot}}{c^2}\left [ 1-\frac{\omega_c}{\omega_0}\right],
\label{superfluid}
\end{equation}
where the f-sum rule value, $f_{tot}$, is proportional to the oscillator
strength of the collective mode.
In Fig.~\ref{fig:two} we show the effects of
pinning on the zero-temperature penetration depth.

The dependence of penetration depth on pinning demonstrates the
importance of distinquishing the macroscopic penetration depth of a
vortex lattice, which characterizes its low frequency response to
long-wavelength electromagnetic radiation, from the local effective
penetration depth\cite{yongwang,affleck} measured by $\mu$SR or NMR
experiments\cite{musrexpt}.  The latter is actually an equilibrium
property of the vortex lattice, and an index of the inhomogeneity of
the internal magnetic field distribution.  (In London theory, the
field distribution is determined by the zero-field penetration depth.)
The $\mu$SR penetration depth is only indirectly related to the
low-frequency limit of the vortex-lattice conductivity.  The
distinction is most stark when disorder and both extrinsic and
intrinsic pinning pinning are absent; the macroscopic penetration
depth should then diverge, while the local $\mu$SR penetration depth
will stay close\cite{yongwang,affleck,longpaper} to its zero-field
value.

Finally, we address the case of inhomogeneous pinning, which we expect
to be be typical for experimental systems and to differ importantly
from the homogeneous pinning case typically studied in theoretical
models\cite{gr,cc}.  Since our magnetic unit cells contain two
vortices, we have the option of pinning them unequally.
Interestingly, we find that the conductivity spectrum for the case of
one pinned vortex per unit cell (bottom left panel,
Fig.~\ref{fig:two}) is not congruent with 
the naive anticipation of independent contributions 
from pinned and free vortices.
Instead, the collective resonance of the  
unpinned and homogeneously pinned cases is 
split, with one peak lying at $\omega <
\omega_c$ and the other at $\omega_c < \omega <
\omega_{\mbox{\scriptsize qp}}$.  If the response were local, a 
peak would occur near $\omega_c$ (perhaps blue-shifted by
inter-vortex forces) and a second smaller peak due to the pinned
vortex would appear near $\omega_{\mbox{\scriptsize qp}}$.
The breakdown of the naive
picture reflects the highly nonlocal nature of the clean vortex
lattice response.

In summary we report on a practical numerical approach for the study of
equilibrium and linear response properties in the mixed state. 
This work is motivated by the growing realization that high $T_c$ 
superconductors are in the clean, long quasiparticle
mean-free-path limit at low temperatures and by the   
need to avoid simplifying approximations which become dubious in this limit.
In this paper we have examined the influence of pinning arrangements 
on the ac conductivity of dense vortex states, demonstrating the 
dominantly collective nature of the response and the essential 
role of non-locality.  This type of calculation can be used to 
test potential paradigms and to help develop sound phenomenologies 
for linear response properties of these complex states.

The authors acknowledge helpful interactions with Dennis Drew, Mohit
Randeria, Ximenes Resende, 
James Sauls, Matthias Eschrig and David Tanner.  This work
was supported by the Midwest Superconductivity Consortium through
D.O.E.  grant no. DE-FG-02-90ER45427, and by the Natural Sciences and 
Engineering Research Council of Canada.


\begin{references}
\bibitem{block} Supercell periodicity permits the mean-field
Hamiltonian to be block diagonalized with block dimensions equal to
twice the number of sites per supercell.  The Bloch transformation
used for this purpose, and explicit expressions for matrix elements in
this representation, will be detailed in Ref. \onlinecite{longpaper}.
\bibitem{hartreecaveat} To simplify the current discussion we have not
included the Hartree and exchange mean fields, which appear in the
diagonal blocks of $H^{BG}$ in either the equilibrium or the linear
response calculation.  Exchange terms treated at the mean-field level
tend to incorrectly yield equlibrium states with diagonal broken
symmetries and are best neglected in model calculations for a
superconductor.  If we include a long-range Coulomb contribution to
the model interaction, the Hartree terms do play a minor role that
will be discussed in Ref. \onlinecite{longpaper}.
\end{references}
\end{document}